%% file: paper.tex
\documentclass{egpubl}
\usepackage{eurova2019}

 \WsPaper           % uncomment for final version of EG Workshop contribution
 \electronicVersion % can be used both for the printed and electronic version

\ifpdf \usepackage[pdftex]{graphicx} \pdfcompresslevel=9
\else \usepackage[dvips]{graphicx} \fi

\PrintedOrElectronic

\usepackage{t1enc,dfadobe}

\usepackage{egweblnk}
\usepackage{cite}

\usepackage{color}

\usepackage{mathtools}
\usepackage{subcaption}
\newlength{\twosubht}
\newsavebox{\twosubbox}
\usepackage{tabularx}
\usepackage{relsize}
\usepackage{xcolor}
\usepackage[export]{adjustbox}
\usepackage{wrapfig}

\usepackage{paralist}

\usepackage{microtype}

\usepackage{soul}

\definecolor{light-gray}{gray}{0.95}

\title[Visual Analytics of Conversational Dynamics]%
{Visual Analytics of Conversational Dynamics}

\author[D. Seebacher, M. T. Fischer et al.]
{\parbox{\textwidth}{\centering Daniel Seebacher$^1$\orcid{0000-0003-0097-5855}, Maximilian T. Fischer$^{1}$\orcid{0000-0001-8076-1376}, Rita Sevastjanova$^{1}$\orcid{0000-0002-2629-9579}, Daniel A. Keim$^{1}$\orcid{0000-0001-7966-9740}, and Mennatallah El-Assady$^{1}$\orcid{0000-0001-8526-2613} 
	}
	\\
	{\parbox{\textwidth}{\centering $^1$University of Konstanz,  Data Analysis and Visualization, Germany \\
		}
	}
}

\begin{document}
	
	 \teaser{
		\vspace{-20pt}
		\includegraphics[width=0.85\linewidth,cfbox=light-gray 1pt 0pt]{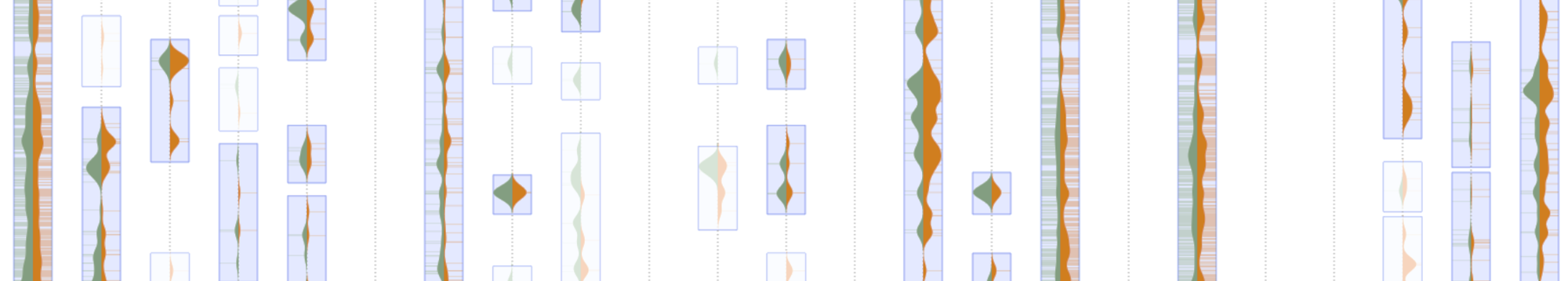}
		\centering
		\caption{Communication Sequence Visualization showing filtered communication {\color[RGB]{111,122,252} \textbf{episodes}} on a vertical timeline. {\color[RGB]{134,180,131} \textbf{Incoming}} and {\color[RGB]{233,125,13} \textbf{outgoing}} communication intensity is shown as a density distribution. In this case, episodes showing a strong \textit{challenge-response pattern} are highlighted.
		}
		\label{fig:teaser}
	}
	
	\maketitle
	
	\input{sections/abstract.tex}
	
    \clubpenalty10000
    \widowpenalty10000
    \displaywidowpenalty=10000 
    \looseness=-1 
    \linepenalty=1000
	
	\input{sections/introduction.tex}
	\input{sections/methodology.tex}
	\input{sections/system.tex}

	\input{sections/discussion.tex}

	\section*{Acknowledgement}
	This project has received funding from the European Union's Horizon 2020 research and innovation programme under grant agreement No 700381 (ASGARD).

	\bibliographystyle{eg-alpha}
	\bibliography{bibliography}

\end{document}

%% file: sections/abstract.tex
% !TeX spellcheck = en_US
%-------------------------------------------------------------------------
\begin{abstract}
Large-scale interaction networks of human communication are often modeled as complex graph structures, obscuring temporal patterns within individual conversations. To facilitate the understanding of such conversational dynamics, episodes with low or high communication activity as well as breaks in communication need to be detected to enable the identification of temporal interaction patterns. Traditional episode detection approaches are highly dependent on the choice of parameters, such as window-size or binning-resolution. In this paper, we present a novel technique for the identification of relevant episodes in bi-directional interaction sequences from abstract communication networks.
We model communication as a continuous density function, allowing for a more robust segmentation into individual episodes and estimation of communication volume. Additionally, we define a tailored feature set to characterize conversational dynamics and enable a user-steered classification of communication behavior. We apply our technique to a real-world corpus of email data from a large European research institution. 
The results show that our technique allows users to effectively define, identify, and analyze relevant communication episodes. 

% ----------------------------------------------------------------------

%  ACM CCS 2012
	% (see http://www.acm.org/about/class/class/2012)
	%The tool at \url{http://dl.acm.org/ccs.cfm} can be used to generate
	% CCS codes.
	%Example:
	\begin{CCSXML}
		<ccs2012>
		<concept>
		<concept_id>10002951.10003227.10003351</concept_id>
		<concept_desc>Information systems~Data mining</concept_desc>
		<concept_significance>500</concept_significance>
		</concept>
		<concept>
		<concept_id>10003033.10003083</concept_id>
		<concept_desc>Networks~Network properties</concept_desc>
		<concept_significance>500</concept_significance>
		</concept>
		<concept>
		<concept_id>10010147.10010178</concept_id>
		<concept_desc>Computing methodologies~Artificial intelligence</concept_desc>
		<concept_significance>500</concept_significance>
		</concept>
		<concept>
		<concept_id>10010405.10010455.10010461</concept_id>
		<concept_desc>Applied computing~Sociology</concept_desc>
		<concept_significance>500</concept_significance>
		</concept>
		</ccs2012>
	\end{CCSXML}
	
	\ccsdesc[500]{Information systems~Data mining}
	\ccsdesc[500]{Networks~Network properties}
	\ccsdesc[500]{Computing methodologies~Artificial intelligence}
	\ccsdesc[500]{Applied computing~Sociology}

% 	\printccsdesc   
\end{abstract}

%% file: sections/introduction.tex
% !TeX spellcheck = en_US
%-------------------------------------------------------------------------
\section{Introduction} \label{sec:introduction}

% intro & motivation
With the digitization of society, especially in our daily %modes of
communication, global information exchange has never been easier, resulting in mounting collections of communication data. The sheer amount% of communication data
, as well as the intertwined structures it is comprised of, pose challenging problems when trying to analyze communication dynamics. Questions such as---what are the patterns underlying the communication network or who are key players?% and who are they?
---are difficult to answer. They not only require the extraction of simple information from these communication datasets, but also the fine-grained analysis of the communication network structure itself to detect patterns in the bi-directional communication behavior between users.

% existing techniques
Addressing these questions, a variety of %computer-based 
approaches were proposed, mainly, with a focus on social network analysis. Examples include the identification of key people in networks or the automatic detection of community structures~\cite{xie_slpa:_2011,xie_overlapping_2013,palowitch_significance-based_2017}. 
In the field of automatic text analysis, text content is examined more closely, for example using sentiment analysis~\cite{pang_opinion_2008}, topic modeling~\cite{el-assady_progressive_2018}, or lexical chaining~\cite{gold-lep-2015}. However, a problem that has not yet received enough attention is %the question of
\emph{how} people communicate with each other, i.e., a detailed exploration of the bi-directional interactions % of entities
within a network. Such analysis allows to draw further conclusions about users' behaviors and relations~\cite{el-assady_contovi:_2016}, thus allowing for more precise identification of roles in social networks.

% our solution
In this paper, we present a novel technique to support experts in their understanding of arbitrary, timestamped interactions, enabling a feature-driven investigation of relevant communication episodes. We use kernel density estimation to model the bi-directional communication events, based on their temporal distribution, as a continuous communication density function. In a second step, we present how to model features based on the communication density and other communication parameters which characterize the bi-directional communication behavior in individual episodes. 

% contributions
Overall, we make the following contributions in this paper. \textit{\textbf{1}}:~A technique for modeling communication based on the temporal distribution of communication events using kernel density estimation. \textit{\textbf{2}}:~Communication density-based detection of communication episodes in bi-directional communication sequences. \textit{\textbf{3}}:~Demonstration of how features can be defined and implemented to characterize the communication behavior in single communication episodes to allow for the visual analysis of those episodes. \textit{\textbf{4:}} A prototype demonstrating the feasibility of this approach as a visual analytics approach for the investigation and analysis of conversational dynamics.

\section{Related Work}

Communication can be seen as social interactions involving numerous entities over time, which leads to large and complex networks. 
The task of analyzing such large networks is generally referred to as social network analysis, which is described %in detail
in the standard literature~\cite{scott_social_2017} and often focuses on using %graph
measures like centrality to analyze social ties and communication behavior~\cite{luo_using_2015}. A general survey of visualization systems for %large
networks is given by Shiravi et al.~\cite{shiravi_survey_2012}. Additionally, since such networks often contain the interactions of millions or billions of entities over time,  simplification is necessary, often using community detection algorithms such as SLPAw~\cite{xie_slpa:_2011} and CCME~\cite{palowitch_significance-based_2017}. An overview of other %possible
techniques is shown in the survey of Aggarwal and Wang~\cite{aggarwal_survey_2010}. 

Approaches that are related to our work and focus on analyzing relations and communications in graph networks include, for example, GestaltMatrix, a matrix-like representation~\cite{brandes_asymmetric_2011}; TimeMatrix, which provides insight about the overall temporal evolution and the activity of nodes over time~\cite{yi_timematrix:_2010}; \textit{Timeline Edges}, which is an integrated approach and tries to leverage unused space in drawing zero-dimensional connectivity information as one-dimensional edges~\cite{reitz_framework_2010}; \textit{T-Cal}, a timeline-based approach that uses distortion to highlight areas with high communication volumes~\cite{fu_t-cal:_2018}, or the methods proposed by Fu et al. recognizing communication patterns~\cite{fu_visualization_2007}. But all of these approaches have drawbacks regarding scaling, comparability, or information overload.

We also employ sequence analysis and, while the task itself is common, most approaches focus exclusively on statistical results or purely on visual comparison~\cite{malik_cohort_2015}. According to Zhao et al.~\cite{zhao_matrixwave:_2015}, only a few have investigated visualization approaches for comparing multiple event sequences. One idea that is proposed is CloudLines~\cite{krstajic_cloudlines:_2011}. Also, a metric has been presented for comparing temporal event sequences, but only for chains of sequences, instead of comparing sequences themselves~\cite{malik_cohort_2015}.

%% file: sections/methodology.tex
% !TeX spellcheck = en_US

\section{Communication Behavior Modeling} \label{sec:methodology}

For the analysis of the communication behavior, we concentrate primarily on the communications between an entity $a$ and another entity $b$, for example, persons or communities. 
The communications between $a$ and $b$ can be considered as the multisets of the edges $(a,b)$ and $(b,a)$ in a communication graph. 
Different questions are of interest when analyzing the communication behavior between these two entities. 
For example, is the volume of communication high or low, is the communication discontinued, and is the communication one-sided (i.e., are there more communications from one entity to the other)? 
To answer such questions for $a,b$, we can compare the number of incoming messages from $b$ with the number of outgoing messages from $a$, or vice versa. 
However, if we look at communications only as individual messages, it may be difficult to answer such questions. 
For example, for finding out if one entity is communicating more than another, we can compare the number of communications at a given time, but this is only possible if communications are compared for the same time ranges. 
If, for example, there is an hour difference between a communication from $a$ to $b$ and the response from $b$ to $a$ (which corresponds to normal response times for e-mails), this would only be measured as a symmetric communication behavior if the communications were also compared on the same time range. 

\begin{figure}%[!ht]
    \captionsetup[subfigure]{aboveskip=-1pt,belowskip=-1pt,justification=justified,singlelinecheck=false}
	\centering
    \begin{subfigure}[b]{\columnwidth}
        \includegraphics[width=\textwidth]{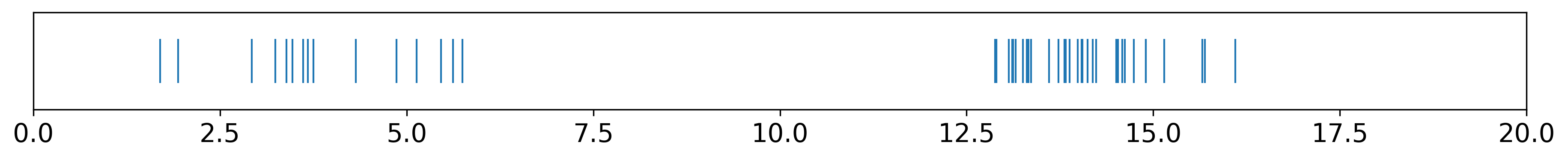}
        \caption{Distribution of communications, on the example of outgoing messages.}
        \label{fig:eventplot}
    \end{subfigure}
    \begin{subfigure}[b]{\columnwidth}
        \includegraphics[width=\textwidth]{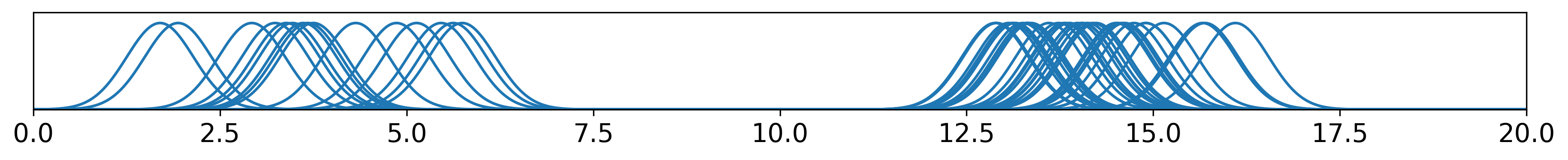}
        \caption{A Gaussian kernel is placed on each communication event.}
        \label{fig:individual_gaussians}
    \end{subfigure}
    \begin{subfigure}[b]{\columnwidth}
        \includegraphics[width=\textwidth]{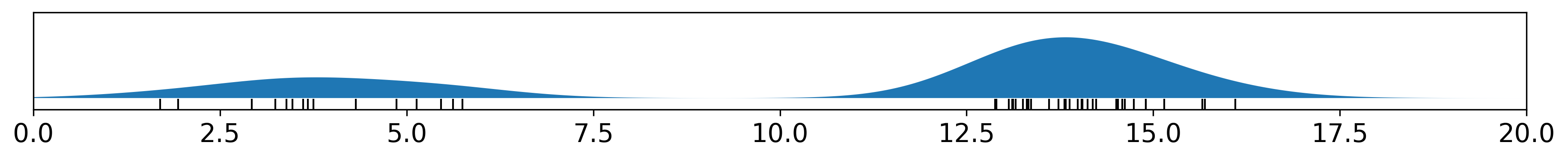}
        \caption{Estimation of the Communication Density using KDE.}
        \label{fig:kde}
        \vspace{-5pt}
    \end{subfigure}	
    \caption{
    Individual communication events are represented as a communication density using KDE. The resulting continuous representation enables a robust detection of communication episodes, as well as, the derivation of features for a classification of such episodes.}
	\label{fig:kerneldensityestimation}
	\vspace{-15pt}
\end{figure}

In order to avoid these problems in the analysis of communication behavior, we do not model the communications as individual events, as shown in \autoref{fig:eventplot}, but as a continuous communication density function, as shown in \autoref{fig:kde}. This avoids the issues with binning or sliding window approaches as described above by using a smooth kernel.
In turn, this prevents problems such as the failed comparison of communication behavior described above, since a communication no longer corresponds to a temporally atomic event, but can be measured with decreasing importance in the past and future and therefore no exact correspondence of the time units must exist anymore.
In order to maintain this continuous communication density function, we use the well-known concept of KDE. 

We replace every communication event between $a$ and $b$ by a Gaussian normal distribution $G(x) = \frac{1}{\sigma \sqrt{2\pi}}e^{-\frac{1}{2}(\frac{x-\mu}{\sigma})^2}$ as shown in \autoref{fig:individual_gaussians}, with $\mu$ being the center of the Gaussian kernel, i.e. the position of the communication event and $\sigma$ the variance.
We can then estimate the communication density $\hat{f}$ for each time point $x$ between the two entities, using the KDE $\hat{f}(x) = \frac{1}{nh}\sum_{i=1}^n G(\frac{x-x_i}{h})$, with $h > 0$ as a smoothing parameter (bandwidth).
The parameters $\mu, \sigma,$ and $h$ can be adjusted as required to make this approach suitable for different domains and tasks. The center $\mu$ is often set to zero (influence exactly around the event time), but could be used to encode a prior or subsequent response. The parameter $\sigma$ describes the temporal influence an individual event has, where a very low value encodes a local event like the existence, whereas higher values could be used to encode more far-reaching concepts like a conversion about a specific topic, which continues for some time. The bandwidth parameter $h$ describes how much individual responses likes spikes should be retained, e.g. for occurrence of key words, or smoothed, e.g. for general tendencies.
If we now consider the communications between two entities $a$ and $b$, we can determine the communication density of the incoming messages $\hat{f}_{in}(a,b)$ (messages from $b$ to $a$) and vice-versa the outgoing messages $\hat{f}_{out}(a,b)$.

By modeling communications as a continuous density function rather than as single atomic communication events, we can easily discover periods with a low or high communication density.
For this, we can directly use the density functions $f_{in}$ and $f_{out}$ to judge whether one or both entities have made several communications in a given period of time.
A further advantage of this approach is that it enables automatic detection of breaks in the communication (i.e., we can conversely identify individual communication episodes). 
For instance, very few people will continually send each other messages over long periods of time.
Much more common is the pattern where one person sends a message that, in turn, leads to a discussion that ultimately ends after a few messages. 
We can determine these individual communication episodes by determining the periods $s$ in which the communication density is greater than a threshold value.
Finally, to enable manual filtering of individual communication episodes as well as visual analysis, we demonstrate how a number of descriptive features for the analysis of communication episodes can be defined. With the help of additional variables such as the length $L_{s_i}$ of one communication episode $s_i$ and the density function for the incoming and the outgoing messages in this communication episode $\hat{f}_{out}^{s_i}$ and $\hat{f}_{in}^{s_i}$, we can then define features which are suitable for manual filtering and also enable a visual analysis of communication behavior of individual communication episodes. An example of such a feature would be synchronicity, i.e., if both entities are involved in a communication to the same extent at the same time.
This would be illustrated by an equal communication density of incoming and outgoing messages in a communication episode.
We can calculate this, for example, by determining the integral of the absolute difference between the two communication densities.

%% file: sections/system.tex
\section{Visual Analytics of Conversational Dynamics} \label{sec:investigation}

In the following section, we want to demonstrate how our technique, in combination with an experimental set of 14 descriptive features, facilitates visual analytics of conversational dynamics. As an example for a real-world dataset, we use email data from a large European research institution~\cite{paranjape_motifs_2017}. The dataset is provided by the Stanford Network Analysis Project and contains the communication of 986 entities over a timespan of 803 days. In total there are 332,334 messages between 24,929 members of the institution.

Using communication density, we present a communication sequence visualization that enables identification of regions with low or high communication behavior. This communication sequence visualization also highlights the individual communication episodes. Finally, we introduce an interactive component that allows the user to manually filter the episodes as well as label existing episodes in order to perform a semi-automatic classification of the communication episodes into user-defined classes.

In order to look at the conversational dynamics in detail, we need to inspect the temporal patterns of incoming and outgoing messages more closely. To help with this, we have developed a visualization of the communication sequences between entities. To represent this conversational dynamic, we can use the communication density $\hat{f}$, defined above. We plot the density of incoming  and outgoing communications $\hat{f}_{in}$ and $\hat{f}_{out}$ as area charts on different sides of a time axis, as shown in \autoref{fig:teaser}.
For the visualization of the density of incoming and outgoing communications, we have selected the subdued colors lime-green and orange and optimized their contrast ratio. In addition, we can also use the communication densities to segment the communication into individual communication episodes by checking whether the density is above a certain threshold $\hat{f}_{in} + \hat{f}_{out} > \varepsilon$. These individual communication episodes are highlighted to make them more distinct, for example with a light blue background. In order to visualize the conversational dynamics amongst multiple users, the individual communication sequences can be arranged side by side. In general, two arrangements are possible: (1) Vertical layout of the communication sequences, as shown in \autoref{fig:teaser}, in order to leverage the width of the display to maximize the number of communication sequences shown. (2) Horizontal layout to leverage the width of the display to maximize the length of the shown communication sequences.

\begin{figure}[b]
\vspace{-10pt}
    % preliminary
    \sbox\twosubbox{%
      \resizebox{\dimexpr0.99\columnwidth-1em}{!}{%
        \includegraphics[height=6cm]{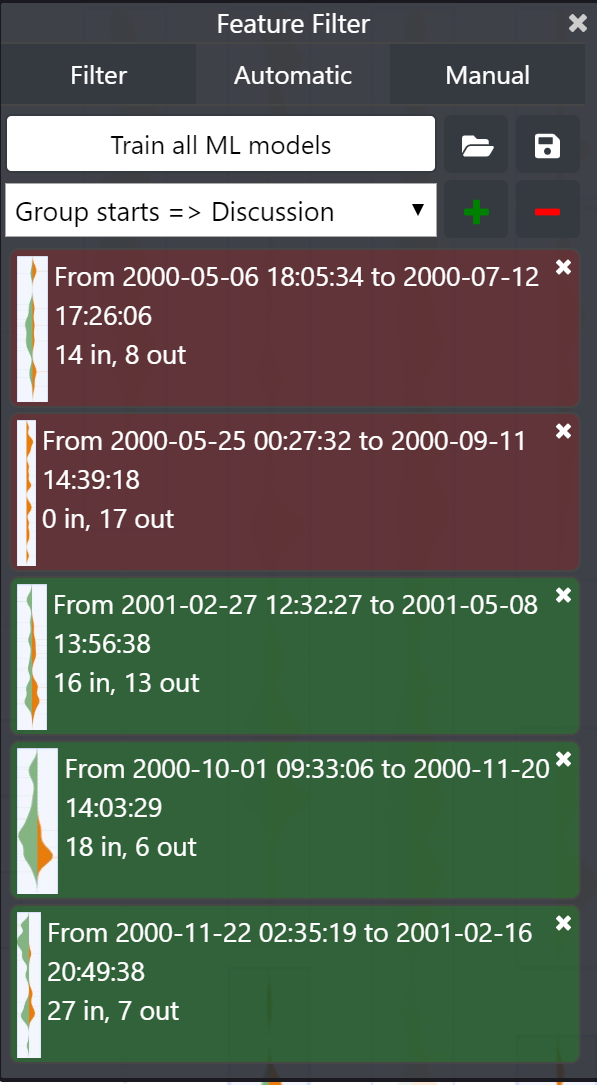}%
        \includegraphics[height=6cm]{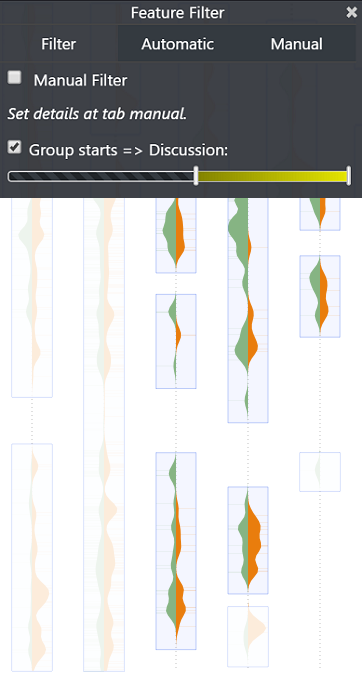}%
      }%
    }
    \setlength{\twosubht}{\ht\twosubbox}
    
    % typeset
    
    \centering
    
    \subcaptionbox{Using {\color[RGB]{50, 98, 57} \textbf{positive}} and {\color[RGB]{125, 51, 56} \textbf{negative}} samples, a ML model is trained to identify episodes in which the selected groups start the conversation, leading to a discussion of both entities.\label{fig:machine-learning}}{%
      \includegraphics[height=\twosubht]{figures/system/ml.png}%
    }\quad
    \subcaptionbox{Application of the trained model to the data. In this example only {\color[RGB]{111,122,252} \textbf{relevant episodes}} with high certainty are displayed, while irrelevant episodes are faded out. \label{fig:filtered-episodes}}{%
      \includegraphics[height=\twosubht]{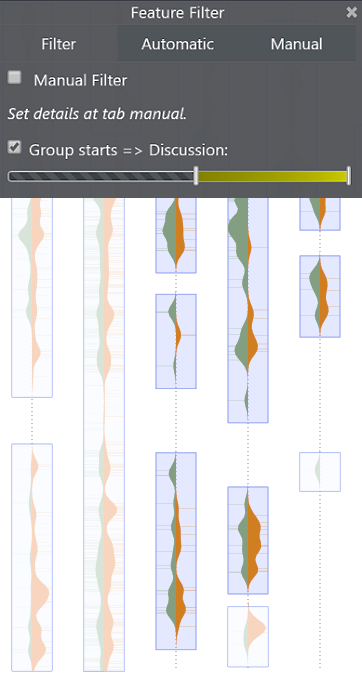}
    }
    \vspace{-5pt}
    \caption{By providing feedback for some data samples, users train ML models to identify relevant conversational dynamics in episodes. }
    \label{fig:machine_learning}
    \vspace{-15pt}
\end{figure}

The concept of communication episodes also differs in their semantic relations, depending on the period under consideration. Communication encompassing several years has to be evaluated differently than one over several days. In the first example, messages may belong to the same episode, even though they might be several days apart. In the second example, however, this would be the entire monitoring period. It is therefore necessary to describe the high-level abstraction of communication differently, depending on the time range under consideration. These different concepts of episodes are supported in our interactive visualization by semantic zooming. The available levels of granularity can be described by relative parameters, best adapted depending on the application domain and the specific analysis task, as described before.

To further enhance the comparability of the episodes, the concepts of timelines is extended; they can represent threads of time that do not need to be consecutive and can represent any number of time-ranges of an arbitrary length. Different pre-defined ranges like days, months, or, for instance, every Monday are available, while user-defined time periods are also configurable. If more than one linear timeline (the default) is selected, all timelines per group are juxtaposed. This makes it possible to compare the conversation dynamics at the same time in several years, which gives a better insight into recurring or changing communication dynamics. To provide further support, the whole view is interactive and each timeline is reorderable and realignable.

To allow for visual analytics of conversational dynamics, we need to be able to classify communication episodes into different classes. However, \textit{a priori}, there is no predefined set of classes in which to classify the episodes. The desirable classes strongly depend on the domain and the analysis task under consideration. Therefore, we present a semi-interactive visual analytics approach where a user can define their own classes by example. A user can define a class and then provide some positive and negative examples as training data by clicking on relevant or irrelevant episodes. Classification is done using machine learning based on the defined features, which ideally show identifiable differences that reflect the user selection.

In our case, as shown in \autoref{fig:machine_learning}, we use a Random Forest Classifier to make this binary match/no match classification with a confidence estimation since it can be trained with very few training samples. This trained classifier can be used to perform the binary classification for all other episodes, representing one model. It is possible to train several models and to combine them to allow for more advanced patterns. Theoretically, a completely manual approach can also work here, using rule-based classification. However, this becomes too tedious for more complex conversation classes and combinations of features and is therefore not practical. Using the semi-automatic approach, a user can define a class and train an appropriate classifier with only a few interactions. Since we use a Random Forest Classifier, we can model the uncertainty for the prediction of each episode. After a user has trained a classifier for a class, we can use this uncertainty measure to additionally filter the episodes. For example, the user can view relevant episodes for a class by choosing only those for which the classifier is very confident. In turn, this also means that we can inspect all episodes for which the classifier is very uncertain about the prediction. These borderline cases are the most promising for re-labeling by the user in order to iteratively optimize the performance of the classifier. 

\textbf{Expert Feedback -- } To evaluate the usefulness of our approach, we conducted an interview with one domain expert. For this interview, a different, proprietary communication dataset was used, whose characteristics are similar to the dataset presented here. The interview was designed as a combined system evaluation and feedback round. The following paragraph describes not only the key findings and comments by the experts, but also possible areas for improvement: The selection of non-consecutive, parallel timelines for comparability is regarded as useful, as well as the dynamic semantic zooming. Some fear was voiced that the default overflow of communication sequences to the right, to reduce the information density, might be misleading and lead to overlooked results. Therefore, it was recommended to compress the whole visualization on the screen initially--even when the density would be too high to be practical--and therefore require zooming all the time, but not leaving anything offscreen. The automatic detection of sequences with semantic zoom (levels of communication) in combination with filtering sequences and applying machine learning models to it is regarded as a very interesting, novel and realistic approach, which is useful to detect and replicate in other timelines or comparing between users. Both the manual filtering as well as the example-based machine learning are judged to be relevant, the former for first exploration and the later for comparison and detection. With these tools, the expert were able to semi-automatically find related patterns, which would be impractical manually. 

In general, the expert interview showed the system works and that the approaches were received with interest and judged to be useful. According to the experts, the system offers many possibilities for different analysis tasks and is well suited for network exploration in the temporal analysis domain. Examples include the examination of bank transactions, phone records, or e-mails, where it proves very useful in specific situations, like finding relevant nodes. The main criticism voiced by the expert is the tendency for information overload when scaling the approach to show the conversational dynamics between numerous entities as they might occur in large communication networks, which might result in overlooked communication.

%% file: sections/discussion.tex
\section{Discussion and Conclusion} \label{sec:discussion}

To demonstrate its feasibility, we applied our framework to parameters relating around communication density and response and have shown how we can visualize and analyze communication behavior with our modeling. This method, however, can be extended to encompass more complex domain-dependent concepts, for instance, message content or sentiment. 
Apart from manual designed features, one can explore the emerging field of automated feature engineering as pioneered by Kanter and Veeramachaneki~\cite{kanter_deep_2015} and Katz et al.~\cite{katz_explorekit:_2016}.
Including own features enables a far more in-depth investigation of conversational dynamics. Nevertheless, the interview with the expert showed that our approach provides benefits when investigating conversational dynamics.
% further features
%

A challenging step for future work is to investigate how this approach can be used for the analysis of conversations of more than two parties, or how it can be integrated into a social network analysis workflow. A potential idea would be to use the communication episodes between entities, found with the help of our approach and classified as relevant by the user, for the weighting of the connection between the entities in a social network graph. Following our VA approach the user can also influence this weight by filtering non-relevant communication episodes. This weighting can than be used to steer community detection algorithms such as SLPAw or as an input for graph layout algorithms to visualize the social network structure. Thus, with previously presented idea to include further domain-specific concepts, such as message content, community detection or layout algorithms could be further steered for answering questions such as whether discussions about relevant topics have taken place between users.